\documentclass[aps,prl,twocolumn,groupedaddress]{revtex4-1}
\usepackage{geometry}
\usepackage[version=3]{mhchem}
\usepackage{geometry}
\usepackage{amsthm}
\usepackage{amsmath}
\usepackage{amssymb}
\usepackage{graphicx}
\usepackage{verbatim}
\usepackage{hyperref}
\usepackage{gensymb}
\usepackage{bbm}

\begin{document}

\title{\ce{PbTiO3}(001) Capped with ZnO(11\={2}0): An Ab-Initio Study of Effect of Substrate Polarization on Interface Composition and \ce{CO2} Dissociation}

\author{Babatunde O. Alawode}
\affiliation{Massachusetts Institute of Technology, Cambridge, MA 02139}

\author{Alexie M. Kolpak}
\thanks{Corresponding author. \\*
kolpak@mit.edu \\*
Present address: 77 Massachusetts Avenue, rm 3-156, 
Cambridge, MA 02139, USA.}
\affiliation{Massachusetts Institute of Technology, Cambridge, MA 02139}

\date{\today}

\begin{abstract}
Catalytic conversion of \ce{CO2} into useful chemicals is an attractive alternative to expensive physical carbon sequestration methods. However, this approach is challenging because current chemical conversion methods employ high temperatures or pressures, thereby increasing cost and potentially leading to net carbon positive processes. In this paper, we examine the interface properties of  ZnO(11\={2}0)/\ce{PbTiO3} and its surface interaction with \ce{CO2}, CO and O. We show that the stoichiometry of the stable interface is dependent on the substrate polarization and can be controlled by changing the growth conditions. Using a model reaction, we demonstrate that a dynamically tuned catalysis schemes could enable significantly lower-energy approaches for \ce{CO2} conversion.

\end{abstract}


\maketitle



Catalytic conversion of \ce{CO2} into fuels or other materials that can be used on an industrial scale is an attractive  alternative to expensive carbon capture and sequestration (CCS) methods currently under consideration. 
As CCS is only feasible when large quantities of \ce{CO2} are generated, preferably close to a suitable geological formation \cite{Haszeldine25092009}, this option cannot be used to address the emissions sources with the greatest collective impact: vehicles and small industrial plants. In contrast, since \ce{CO2} can be used a precursor for the synthesis of numerous, industrially relevant  carbon based compounds, chemical sequestration approaches could in principle be tailored to smaller scale applications, with implementation costs offset by the production of value-added chemicals.  Although conceptually appealing, chemical conversion approaches are challenging due to the exceptional stability of the \ce{CO2} molecule.  As a result, \ce{CO2} conversion reactions are performed under energetically costly conditions (i.e., high temperature and/or high pressure) that mitigate the effects of chemical sequestration; in fact, this can lead to some reactions being net carbon positive, generating more \ce{CO2} than they consume  \cite{CSSC:CSSC201000447}. In order to make chemical sequestration feasible, new catalysts that can operate under low temperature and pressure conditions need to be developed.

\begin{figure}
\includegraphics[scale=0.30]{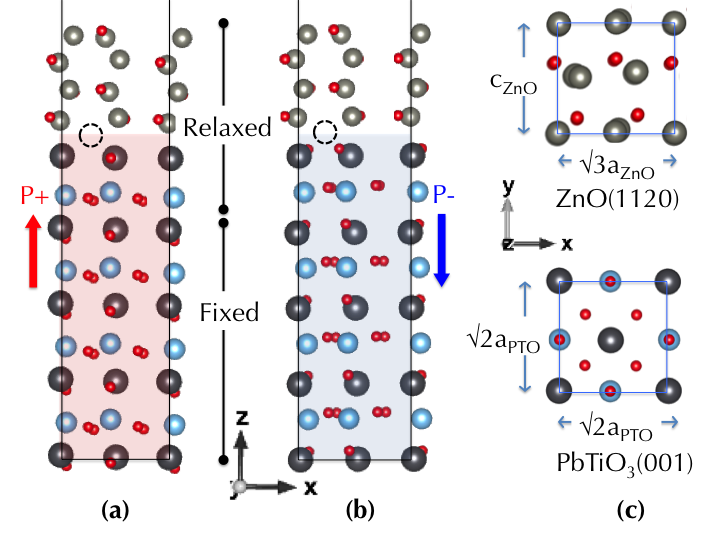}
\caption{Computation supercell and relaxed atomic structure  for (a) ZnO(11\={2}0)$_{n}$/\ce{PbTiO3}$\uparrow$ and (b) ZnO(11\={2}0)$_{n}$/\ce{PbTiO3}$\downarrow$ slabs for $n=4$. The dashed circle represents the position for oxygen insertion at the interface. Fig. (c) shows
the parameters and orientations for matching ZnO(11\={2}0) to \ce{PbTiO3}. Grey, red, cyan, and black atoms are Zn, O, Ti, and Pb, respectively.
\label{fig:Unit-cell-and}}
\end{figure}

Various studies have shown that the polarization of a substrate can affect its surface properties \cite{stadler1965changing,parravano1952ferroelectric}, and this effect has been applied to studies of molecular adsorption \cite{garrity_chemistry_2010,doi:10.1021/acscatal.5b00507,riefer_polarization-dependent_2012,li_direct_2008} and the modulation of the carrier density in conductors \cite{baeumer_ferroelectrically_2015}. Kolpak {\em et al.} suggested the possibility of using dynamical control of surface structure and reactivity in the
coupling of ferroelectric \ce{PbTiO3} with Pt \cite{kolpak2007polarization}.
Performing density functional theory (DFT) computations of  molecular and atomic adsorption to the surface of ultrathin Pt(100) films supported on ferroelectric
\ce{PbTiO3}, they showed that switching the polarization
direction of the substrate dramatically changes the chemisorption strength
and site preference of CO, O, C, and N, potentially altering the reaction
pathways for dissociation of CO, \ce{O2}, \ce{N2}
and NO. However,  polarization-induced changes in the surface chemistry effects are mitigated by the small electronic screening length of Pt, so that only atomically thick films exhibit significant effects. Moreover, it is challenging to wet oxide surfaces with catalytically active transition metals \cite{campbell_ultrathin_1997}; thus,  aggregation of the metal into nanoparticles is expected to supress the effects of substrate polarization on catalytic activity. In this work, we attempt to mitigate these challenges
 by  considering the use of a ferroelectric substrate to tune the surface properties of thin films of an insulating catalytic oxide. In particular, we use DFT to investigate the surface chemistry of thin ZnO(11\={2}0) films supported on ferroelectric \ce{PbTiO3}.

Zinc oxide is commonly used as a catalyst for the industrially and
environmentally important CO and \ce{CO2} conversion
reactions, frequently in conjunction with a copper co-catalyst \cite{Behrens18052012,Grunwaldt2000452,ANGE:ANGE200702600,PhysRevB.47.13782}. Therefore the properties of the various ZnO surface
terminations have been extensively studied \cite{kurtz2005activesites,wang1987surface,meyer2003densityfunctional,kossmann2012prediction}.  We choose to study the epitaxial interface of non-polar
ZnO(11\={2}0) films with \ce{PbTiO3} (PTO), as similar systems have previously been grown. Wei {\em et. al.}
\cite{wei2007heteroepitaxial} reported heteroepitaxial
growth of ZnO(11\={2}0) on \ce{SrTiO3}(001) and
\ce{BaTiO3}(001)/\ce{SrTiO3}(001) surfaces,
suggesting that \ce{PbTiO3}(001), which has the same crystal structure
and very similar in-plane lattice constant, will also provide an
experimentally feasible substrate for ZnO(11\={2}0) films.

To investigate the ZnO(11\={2}0)/\ce{PbTiO3}(001) heterostructure, we perform DFT computations using the plane-wave pseudopotential code
Quantum Espresso \cite{paologiannozzi2009quantum} with ultrasoft
pseudopotentials \cite{vanderbilt1990softselfconsistent} and a 35Ry energy cutoff. The
Wu-Cohen GGA functional \cite{wu2006moreaccurate} is used
to describe exchange correlation; this functional has been shown to
have equivalent or better performance to the ubiquitous PBE GGA functional for the
prediction of structural and energetic properties of ferroelectric
perovskite oxides, as well as a range of other solids, surfaces, and
molecules \cite{tran2007performance}.

To model the heterostructure, we use the experimentally reported
ZnO(11\={2}0)/\ce{SrTiO3}(001) epitaxial relationship
\cite{wei2007heteroepitaxial}. The
PbO-termination of the \ce{PbTiO3}(001) slabs is selected, as this has
been demonstrated to be the thermodynamically favored slab termination
under relevant conditions \cite{garrity2013ferroelectric}. Below, we
use the notation (ZnO)$_{n}$/\ce{PbTiO3}$\uparrow$
and (ZnO)$_{n}$/\ce{PbTiO3}$\downarrow$ for ZnO(11\={2}0) grown on
positively polarized (``up'') and negatively polarized (``down'')
\ce{PbTiO3}(001) slabs, respectively, where $n$ is the
number of ZnO(11\={2}0) atomic layers.  The supercell geometry is illustrated in
Fig. \ref{fig:Unit-cell-and}. A PbO-terminated  cell
with nine alternating PbO and Ti\ce{O2} atomic layers
stacked in the (001) direction is used to represent the ferroelectric
\ce{PbTiO3} substrate. For calculations without \ce{CO2} adsorption or dissociation, we use a $c$(2$\times$2) PTO cell for which a  $4\times4\times1$ $k$-point
mesh is sufficient. For other calculations, we use a (2$\times$2) PTO cell. A 20\AA  \enspace vacuum was added between periodic images in the $z$-direction
and a dipole correction
\cite{bengtsson1999dipolecorrection} is applied in the center of the
vacuum region to remove artificial fields between periodic images in all calculations. 

\begin{figure}
\includegraphics[scale=0.35]{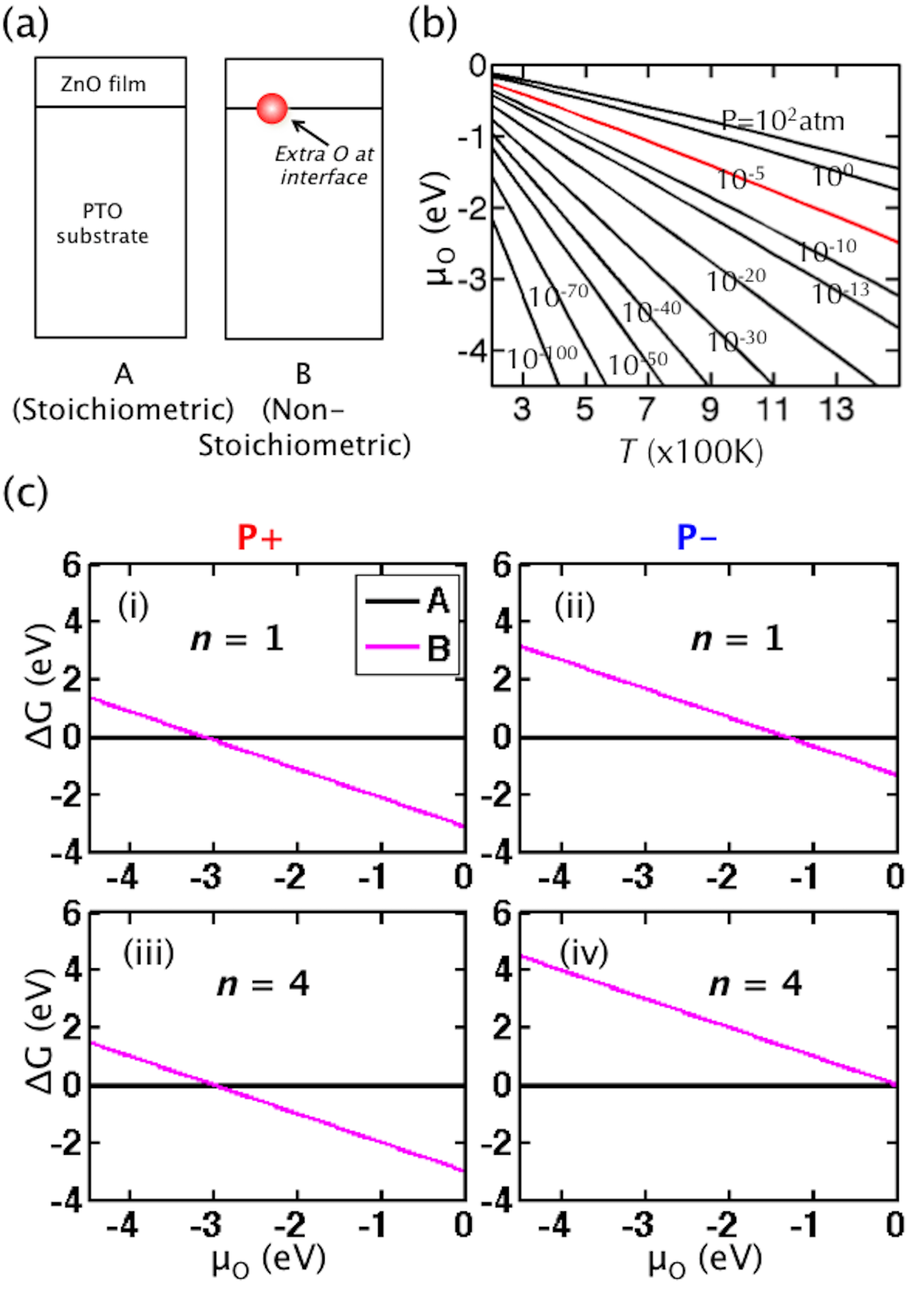}
\caption{(a) Possible configurations in an oxygen-rich environment. (b) Oxygen chemical potential as a function of temperature and pressure. (c) Free energy versus oxygen chemical potential for $n=1$ and $n=4$. shaded regions represent the range of $\mu_O$ under the growth conditions reported in Ref. \cite{wei2007heteroepitaxial} is $10^{-5}$Pa (highlighted line) and 400$\degree$C $< T <$ 700$\degree$. $\mu_O$ can be decreased by reducing pressure and increasing temperature.
 \label{fig:free-energy-plots}}
\end{figure}

We first investigate  possible thermodynamic ground states of the heterostructure under excess oxygen as would be expected during growth and device operation. Figure \ref{fig:free-energy-plots} is a plot of the free energy of formation for inserting an oxygen into the space created at the interface by the unmatched stacking of atomic layers. Our results show that under growth regular conditions (T=400-700\degree C and P=$10^{-5}$Pa \cite{wei2007heteroepitaxial}  at which $\mu_O \approx -1.1 - -1.5$eV) an extra oxygen will remain at the interface when the  subtrate is positively polarized. On the negatively polarized structure, this non-stoichiometric structure is only stable at lower temperatures.

\begin{figure}
\includegraphics[scale=0.36]{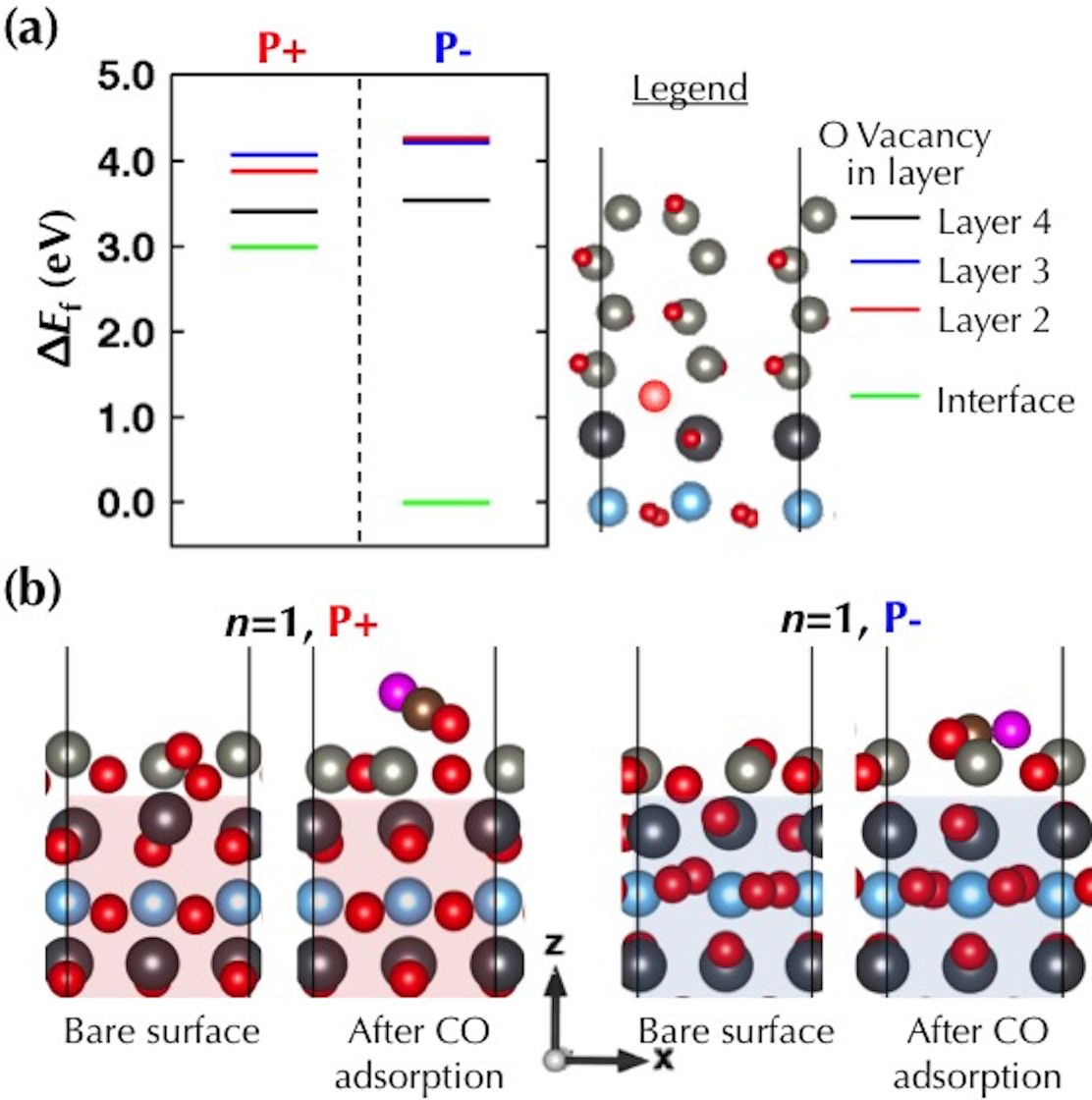}
\caption{(a) Oxygen vacancy formation energy,  $\Delta E_f$, as a function of distance from the interface. The presence of an extra O at the interface is predicted to be determined by the initial growth configuration. (b) Atomic structure of  CO on the non-stoichiometric (ZnO)$_{1}$/\ce{PbTiO3}$\uparrow$ and (ZnO)$_{1}$/\ce{PbTiO3}$\downarrow$, demonstrating the selective removal of interfacial O by the adsorbing CO.  Atom colors are the same as in Fig. \ref{fig:Unit-cell-and}, with adsorbate O atoms shown in magenta for clarity.
 \label{fig:vacancy}}
\end{figure}

In order to decide whether or not the structure with an extra oxygen is relevant for our \ce{CO2} dissociation calculations, we perform calculations for the energy required to have an oxygen vacancy in each ZnO layer in the 4-layer-thick supported-ZnO case. Figure \ref{fig:vacancy}(a), which shows the O vacancy formation energy in each layer for this film, suggests that after a number of layers have been grown, the configuration at the interface will be maintained irrespective of the thermodynamic stability of the structure. If an oxygen is already trapped at the interface (for example, if the film is grown over a positively polarized substrate and the polarization was later switched), its removal will involve a series of steps that include the removal of one of the topmost oxygen atoms. Given the high vacancy formation energies of $\sim$3.5eV, this process will be kinetically limited. Conversely, if there is no oxygen at the interface (for example, growing the under a negatively polarized substrate and the polarization was later switched), oxygen insertion will involve one of the oxygen atoms in the lowest ZnO layer moving to the interfacial O location and leaving behind a vacancy. In our calculations, we do not observe a stable system with this configuration.  

However, the above considerations do not take into account the effect of other gas phase molecules in the environment. For \ce{CO2} dissociation, for example, CO adsorption also occurs. Our calculations show that if an oxygen atom exists at the interface, it can be easily removed by adsorption of a CO molecule, as illustrated in figure \ref{fig:vacancy}(b) for $n=1$ and 2. This suggests an interesting application of non-stoichiometric (ZnO)$_{n}$/\ce{PbTiO3}$\uparrow$ for small $n$: catalytic oxidation of CO to \ce{CO2}, since interface stability dictates that the non-stoichiometric interface will exist as long as there is \ce{O2} gas in the atmosphere at most conditions encountered in practice, and this structure easily oxidizes CO.

The foregoing also suggests that it is possible to obtain a stoichiometric interface if the ZnO thin film is grown at the negatively-polarized structure under conditions of lower oxygen partial pressure and/or higher temperature (i.e., lower oxygen chemical potential, $\mu_O$). This configuration will be trapped when the polarization of the substrate is switched with an electric field. In the following, we perform calculations to explore \ce{CO2} dissociation on the surfaces of this structure as a function of polarization direction and ZnO film thickness. We then show that a dynamically tunable scheme using this heterostructure can enable a much lower overall activation energy for the \ce{CO2} dissociation process.

\begin{figure}
\includegraphics[scale=0.32]{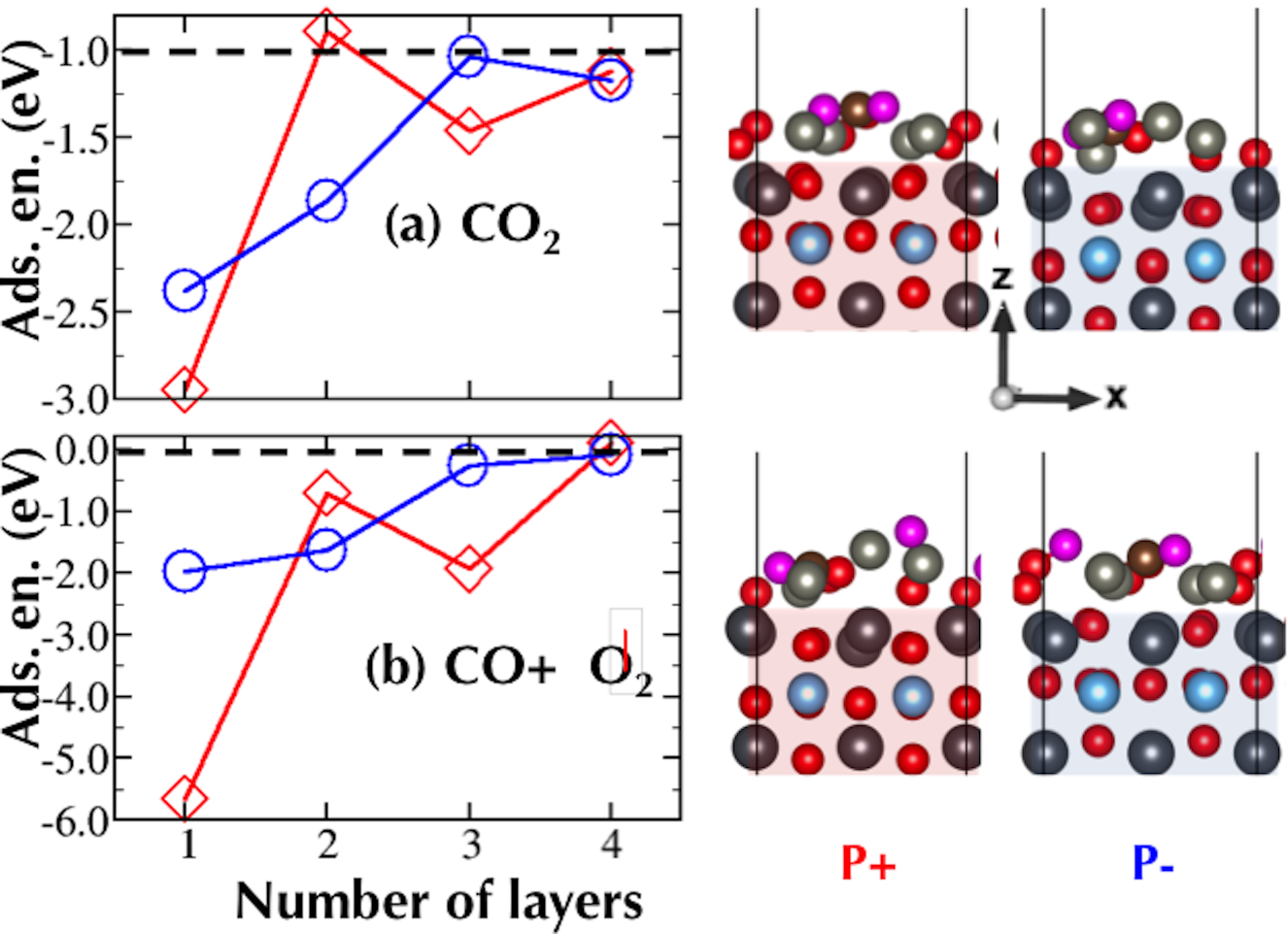}
\caption{Adsorption energies of (a) \ce{CO2}, and (b) \ce{CO}+$\frac{1}{2}$\ce{O2}, on (ZnO(11\={2}0))$_n$/[$2\times2$]\ce{PbTiO3} as a function of $n$. Diamonds and circles represent adsorption on the positively and negatively polarized structures, respectively. horizontal dashed lines represent the adsorption energy of the molecules on an unsupported ZnO slab. Structures on the right  show the adsorption configurations for the corresponding molecule on stoichiometric structures with $n=$1. Oxygen atoms in the adsorbates are colored magenta for clarity.
 \label{fig:Adsorption-properties-of}}
\end{figure}

We perform calculation of the adsorption energies of \ce{CO2} and CO+$\frac{1}{2}$\ce{CO2}  on the surfaces using a coverage of one molecular adsorbate per
2$\times$2 \ce{PbTiO3} surface cell, starting from a range of initial
adsorbate positions and geometries.  The reported adsorption energies in Fig. \ref{fig:Adsorption-properties-of}(a) are
for the minimum energy adsorption geometry on the indicated
surface.  As the figure
demonstrates, changing the polarization direction of the substrate
leads to a difference of $\sim0.6$eV in the \ce{CO2} adsorption
energy for the thinnest ZnO film. The results demonstrate the changing 
surface chemistry as a function of substrate polarization and film thickness. The difference tapes off, with the heterostructure behaving very similarly to the unsupported ZnO slab
for $n\geqslant4$ ZnO layers. Similar behavior is seen for CO+$\frac{1}{2}$\ce{CO2}, as shown in Fig. \ref{fig:Adsorption-properties-of}(b). 

The flip-flop pattern in the adsorption energies in the plots in Fig. \ref{fig:Adsorption-properties-of} can be directly correlated to the
cation displacements at the slab surface, which has the same form with respect to number of ZnO layers and substrate polarization (see the supplemental material \cite{supplemental}).  We find that the magnitude of the displacement in each ZnO layer depends on the distance from the \ce{PbTiO3} surface and its polarization, becoming approximately the same as that in surface of the unsupported thick ZnO after the third layer for both substrate polarization directions. In other words, the magnitude and degree of decay of the surface zinc-oxygen displacements away from the substrate are directly related to the dipole at the interface. Therefore, by changing the magnitude of the displacement at the interface, it may be possible to modify the displacement at the ZnO surface and thus the sensitivity of the tunable catalyst.
To confirm this hypothesis, we perform calculations using the less stable \cite{PhysRevB.88.045401}  \ce{TiO2}-terminated surface, which exhibits smaller displacements than the PbO termination. We find that  \ce{CO2} adsorption on one-layer of ZnO supported on the \ce{TiO2}-terminated substrate is $\sim$10\% ($\sim$54\%) as strong as that on the positively (negatively) polarized PbO-terminated substrate, suggesting that an important design principle for a tunable catalyst with ZnO is to find subtrates with reversible polarizations and large surface displacements.    We predict, for example, that the ferroelectricity-induced ZnO surface chemistry changes will be greater for \ce{BiMO3} than \ce{BaTiO3} or \ce{PbTiO3} substrates.


\begin{figure}
\includegraphics[scale=0.28]{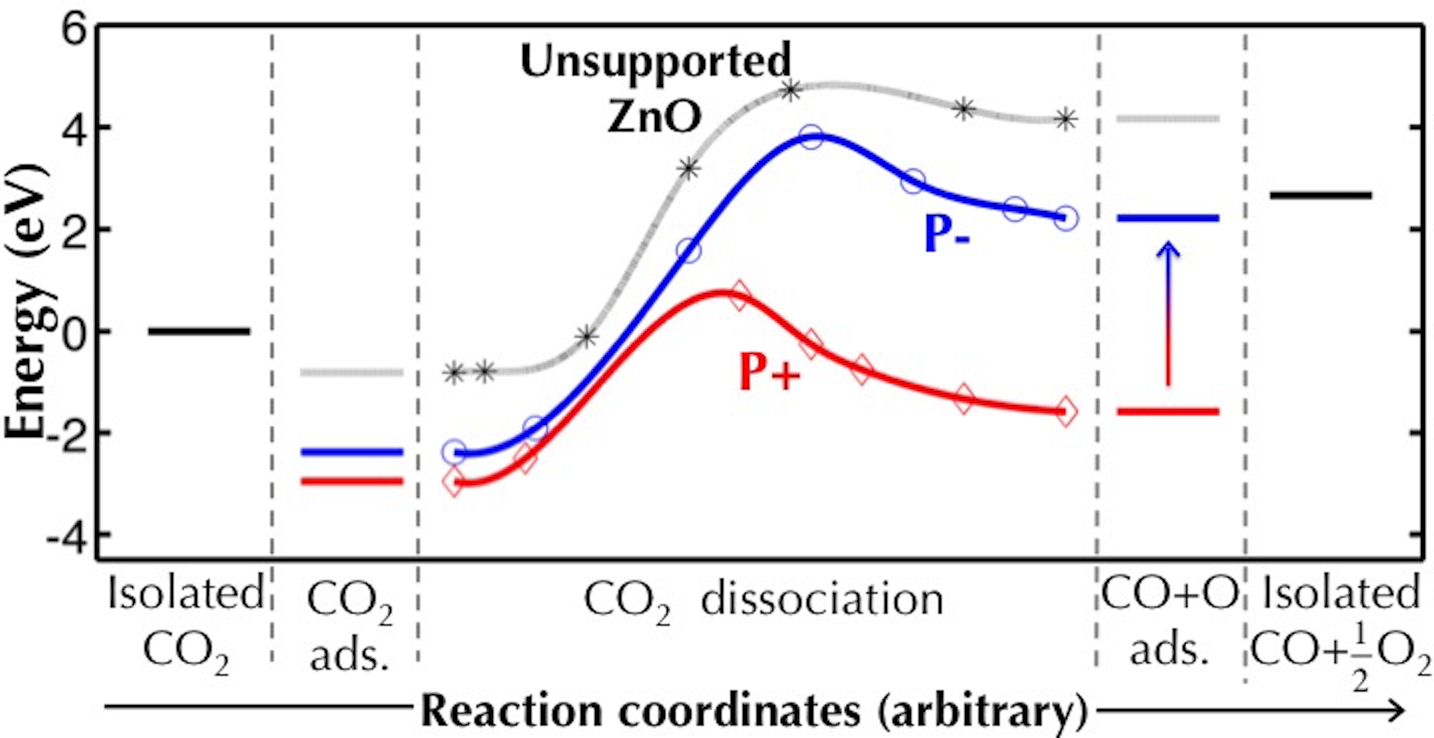}
\caption{Reaction pathway for \ce{CO2} dissociation over unsupported ZnO (gray stars) and ZnO supported on positively and negatively poled \ce{PbTiO3} (red diamonds and blue circles, respectively).   The vertical axis is energy with respect to an isolated \ce{CO2} molecule. The vertical arrow indicates switching of the substrate polarization from P+ to P-.
 \label{fig:pathways}}
\end{figure}

As adsorption energy \cite{adsen1,adsen2}, adsorption geometry \cite{config},
 and surface charges (dipoles) \cite{TiO2ferro} all
affect reaction pathways and energetics, our results strongly suggest
that the polarization-induced changes in these properties will affect
\ce{CO2} conversion processes. 
We investigate this idea further
by determining the reaction pathway and energy barrier for a simple
conversion process: thermally-activated dissociation of \ce{CO2}(g) into
CO(g) and \ce{O2}(g) over a reversibly tunable surfaces of (ZnO)$_n$/\ce{PbTiO3}. To compute the energy along the reaction pathway for each system, we use nudged elastic
band calculations \cite{henkelman2000aclimbing,sheppard2008optimization} to determine the reaction pathways. Figure \ref{fig:pathways} shows the computed reaction
pathways for \ce{CO2}(g) adsorption and conversion to CO(g) and \ce{O2}(g)
over the ferroelectric-supported ZnO films with $n=1$ and on the
unsupported ZnO(11\={2}0) slab. The first step of the reaction,
\ce{CO2} adsorption, is a spontaneous process on all three
surfaces. The dissociation of adsorbed
\ce{CO2} into adsorbed CO and adsorbed O is endothermic for all three
surfaces, but is significantly more favorable on the positively
polarized structure (red line), with an activation energy barrier less
than half that of the reference unsupported ZnO slab (dashed line). The last step, desorption of CO $+$
$\frac{1}{2}$\ce{O2}, however requires a large energy input for
the positively polarized structure, but occurs spontaneously for the
reference surface and requires only minimal energy for the negatively
polarized structure.

Our results suggest a possible approach for increasing the rate of
\ce{CO2} dissociation on ZnO surfaces: grow a ZnO thin film
on a positively polarized substrate, on which the first few steps of
the reaction take place, switch the polarization (depicted by the arrow in the figure) to induce the desorption of
the products, then switch back and repeat. Using a quasi-equilibrium approximation (see supplemental material \cite{supplemental}), our calculations show that
the dynamic switching scheme will result in 10-20 orders of magnitude increase in reaction rates compared to the dissociation on the unsupported slab. Also, at a given rate, the reaction on the tunable catalyst proceeds at less than half of the temperature required for the process on the unsupported slab.  These results clearly demonstrate the potential advantages
of dynamically switching the surface properties of a catalyst to enhance the turnover of a product.

In conclusion, we have shown that the configuration at the interface is dependent on the growth conditions and polarization of the substrate. We also show that the surface chemistry of stoichiometric (ZnO(11\={2}0))$_{n}$/\ce{PbTiO3}
is dependent on both the polarization direction of the \ce{PbTiO3} substrate and the number of ZnO(11\={2}0) layers $n$, with the effect of the substrate polarization becoming negligible for $n\geq4$. The large changes in \ce{CO2} adsorption energy with polarization switching reported in this work suggest the possibility of controlling reaction energetics and pathways of \ce{CO2} reactions, as indicated by the proposed dynamic polarization-switching scheme for a \ce{CO2} dissociation reaction on ZnO(11\={2}0)/\ce{PbTiO3}. Finally, we note that this approach can be applied to many other reactions in heterogenous catalysis, potentially opening new avenues for controlling reaction energetics.

\begin{acknowledgments}
\vspace{5mm}
This work was supported by an MIT Energy Initiative seed grant. We gratefully acknowledge the use of computing resources from NERSC and TACC. We appreciate stimulating discussions with Dr. Brian Kolb about this work.
\end{acknowledgments}

\bibliography{ref.bib}

\end{document}


\title{Supplemental Material for \\ \ce{PbTiO3}(001) Capped with ZnO(11\={2}0): An Ab-Initio Study of Effect of Substrate Polarization on Interface Composition and \ce{CO2} Dissociation}

\author{Babatunde Alawode}
\affiliation{Massachusetts Institute of Technology, Cambridge, MA 02139}

\author{Alexie Kolpak}
\thanks{Corresponding author. \\*
kolpak@mit.edu \\*
Present address: 77 Massachusetts Avenue, rm 3-156, 
Cambridge, MA 02139, USA.}
\affiliation{Massachusetts Institute of Technology, Cambridge, MA 02139}

\maketitle

\section{Effect of adding an electrode}

An electrode or some kind of support
is necessary for most applications
involving perovskites. Electrodes
have been demonstrated to have
significant effects on the perovskite
properties. For example, Sai \emph{et
al} \cite{sai2005ferroelectricity}
reported that the Pt electrodes
cancel 97\% of the depolarizing
field in thin \ce{PbTiO3}
thin films and thus help to maintain
some polarization even in films
one lattice unit thick. The
grounded electrodes provides metallic
screening that compensates the
polarization charge. Arras \emph{et
al.} \cite{arras2012tuningthe}
carried out an interesting study
on the effects of metal electrodes
on the \ce{LaAlO3}/\ce{SrTiO3}
interface. They showed that changing
the type of metal greatly affects
the Schottky barrier, carrier concentration
and lattice polarization at the
interface.

In the light of these, it is imperative
to understand how adding a metal
electrode in our model of the ZnO/\ce{PbTiO3}
affects the ZnO surface (hence
catalytic) properties. If the electrode
has an effect, we will have a better
understanding of phenomena at the
surfaces. If it does not, then
we can get away with modeling a
ZnO/\ce{PbTiO3} system
with fewer atoms hence less computational
costs.

\begin{figure}
    \centering
    \subfloat[Schematic]{\includegraphics[scale=0.3]{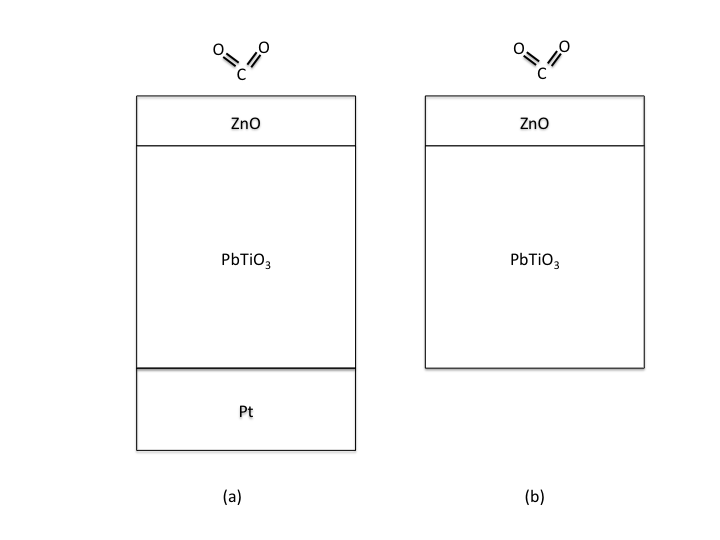}}
    \label{fig:Determining-the-effects}\qquad
    \subfloat[Results]{\includegraphics[scale=0.3]{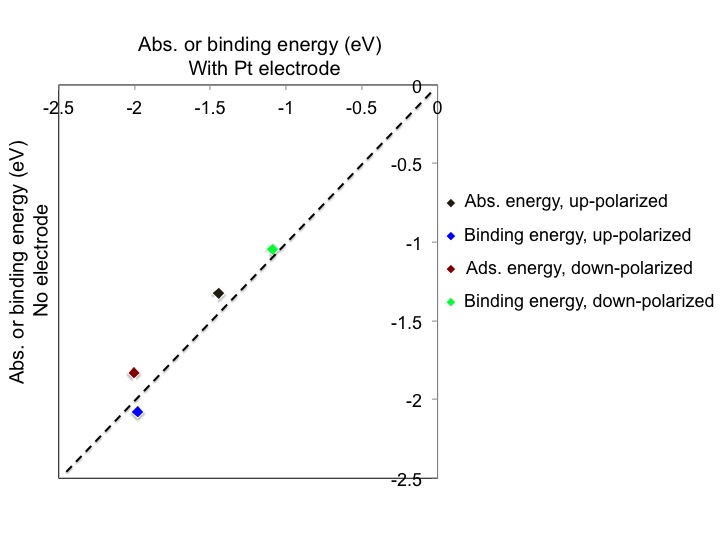}}
\label{fig:CO2-activ}
    \caption{(a) Determining the effects of platinum
electrodes on surface properties.
Calculations were carried out
with an electrode support and
without an electrode support. (b)Comparing the ZnO binding energy
on \ce{PbTiO3} and
\ce{CO2} adsorption
energy on ZnO/\ce{PbTiO3}
with and without a Pt electrode.}
    \label{fig:ComparingtheZnObindingenergy}
\end{figure}

We use two ZnO
layers, four lattice parameter
thick PbO-terminated \ce{PbTiO3}
slab (bottom three fixed) and four
layers of Pt at the bottom. We
calculate the ZnO binding energy
\ce{PbTiO3}/Pt and
\ce{PbTiO3}, density
of states of ZnO/\ce{PbTiO3}/Pt
and ZnO/\ce{PbTiO3},
and \ce{CO2} adsorption
energy on ZnO/\ce{PbTiO3}/Pt
and ZnO/\ce{PbTiO3}. With the Pt and ZnO layers and
the topmost \ce{PbTiO3}layer
relaxed, we calculate the binding
energy of ZnO to \ce{PbTiO3}
in the presence and absence of
the electrode. We find that
the electrode has no effect of
the ZnO binding or \ce{CO2} adsorption. The results are presented in Fig. \ref{fig:ComparingtheZnObindingenergy}.

\begin{figure}
    \centering
    \subfloat[With electrode]{\includegraphics[scale=0.3]{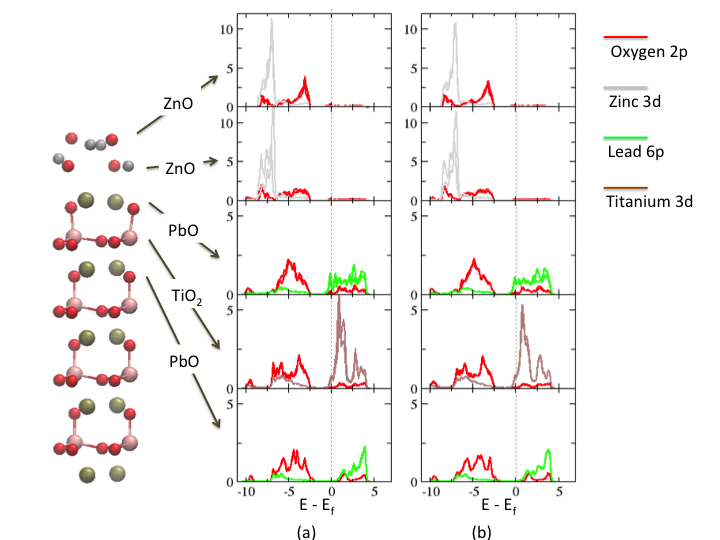}}
    \label{fig:Projected-densities-of1}\qquad
    \subfloat[No electrode]{\includegraphics[scale=0.3]{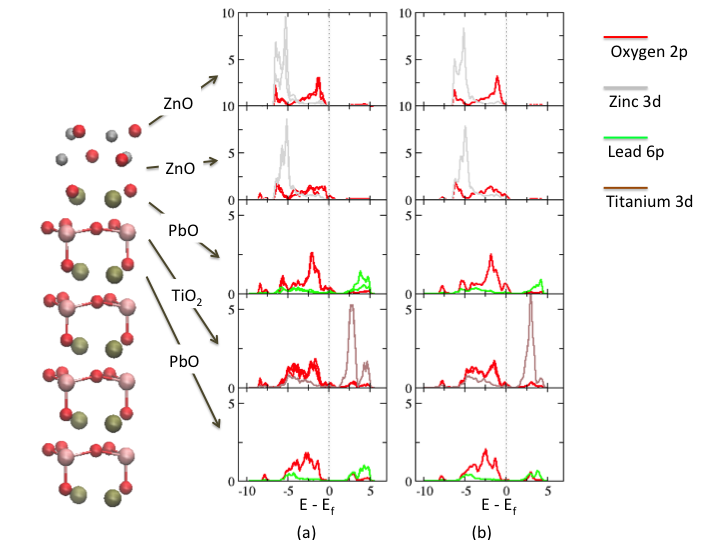}}
\caption{Projected densities of states for
the topmost five layers of a) (ZnO)$_2$/\ce{PbTiO3}/Pt
and b) (ZnO)$_2$/\ce{PbTiO3}.}
\label{fig:Projected-densities-of1}
\end{figure}

Finally, we examine whether an electrode affects the electronic structure of the surface. As seen
in Fig. \ref{fig:Projected-densities-of1},
there is no significant change
in the electronic structure of
the layer both for the up- and
down- polarized catalyst when the
platinum electrode is removed.

From the foregoing, it is evident
the electrode is not important
to consider. Therefore, in
this work, we model ZnO/\ce{PbTiO3}/Pt
as ZnO/\ce{PbTiO3}.

\section{Mechanism for interface-mediated surface chemistry}
\begin{figure}
\includegraphics[scale=0.40]{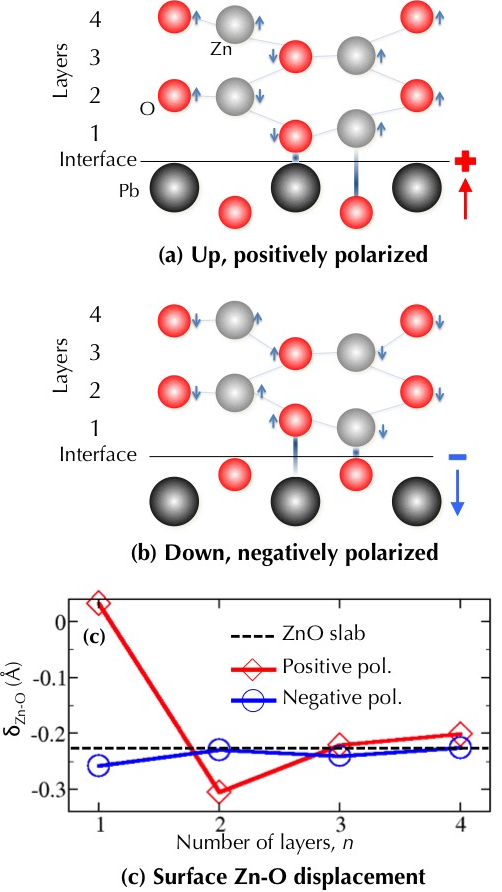}
\caption{Mediation of the interface chemistry by the (a) positively polarized (b) negatively polarized substrates. (c) is a plot of the surface displacements.
 \label{fig:mechanism}}
\end{figure}

An inspection of the relaxed configuration of the heterostructures show changes in the surface Zn-O displacements with respect to the unsupported slab. The mechanism for this is shown illustrated in Fig. \ref{fig:mechanism}(a) and (b). The attraction of the Pb atoms on the substrate side at the interface of the positively-polarized structure to the O atoms in the ZnO film causes an increase in the Zn-O displacement in that layer. The next layer however experiences a decrease in its Zn-O displacement, and so on. The opposite effect explains the trend for the negatively polarized structure. The effect of the interface mediation gets weaker as the surface layers gets farther from the surface. The data plotted Fig. \ref{fig:mechanism}(c) suggests that the interface mediated mechanism is correct.

\section{Rate equation for \ce{CO2} dissociation on ZnO and ZnO/\ce{PbTiO3} }

Step 1 (\ce{CO2} adsorption): \ce{CO2 + 3 * -> CO2 ***}

Step 2 (\ce{CO2} dissociation): \ce{CO2 *** -> CO ** + O *} . This is the rate limiting step 

Step 3 (CO desorption): \ce{CO ** + O * -> CO + 2 * + O *} 

Step 4 (Oxygen desorption): \ce{O * -> 1/2 O2 + *} 

We apply quasi-static approximation (all steps except the rate limiting step are in equilibrium). We define the  rate constants for each step.
\begin{equation} \label{eq:1}
K_1 =\frac{ \theta_\ce{CO2}}{\frac{P_\ce{CO2}}{P_{ref}}{\theta_*}^3}
\end{equation}
\begin{equation} \label{eq:2}
K_2 =\frac{ \theta_\ce{CO}\theta_\ce{O}}{\theta_\ce{CO2}}
\end{equation}
\begin{equation} \label{eq:3}
K_3 =\frac{ {\theta_*}^2\frac{P_\ce{CO}}{P_{ref}}}{\theta_\ce{CO}}
\end{equation}
\begin{equation} \label{eq:4}
K_4 =\frac{ {\theta_*}\left(\frac{P_\ce{O2}}{P_{ref}}\right)^{1/2}}{\theta_\ce{O}}
\end{equation}

The coverages \(\theta\) are related by:
\[ \theta_\ce{CO2} +  \theta_\ce{CO}   +  \theta_\ce{O} +  \theta_* = 1\] which gives

\begin{equation} \label{eq:5}
K_1 \left(\frac{P_\ce{O2}}{P_{ref}}\right){\theta_*}^3   + \frac{1}{K_3} \left( \frac{P_\ce{CO}}{P_{ref}}\right){\theta_*}^2  +   \frac{1}{K_4} \left( \frac{P_\ce{CO}}{P_{ref}}\right)^{1/2}{\theta_*}	+  \theta_* = 1
\end{equation}
We can find the coverages by solving this equation.

The overall rate of reaction, in number of \ce{CO2}converted per second, is given by the rate of the limiting step
\begin{align*} \label{eq:6}
  R_{overall} = R_2 &= k_2\theta_\ce{CO2}   - k_{2-}\theta_\ce{O}\theta_\ce{CO} \\
&= k_2\theta_\ce{CO2}  - \frac{k_{2}}{K_2}\theta_\ce{O}\theta_\ce{CO} \\
&= k_2 \theta_\ce{CO2} - k_{2}\theta_\ce{CO2} \frac{\theta_\ce{O}\theta_\ce{CO}}{K_{2}\theta_\ce{CO2}} \\
&= k_2 \theta_\ce{CO2} (1 - \beta) \\
&= k_2 K_1\left(\frac{P_\ce{CO2}}{P_{ref}}\right){\theta_*}^3(1 - \beta)
\end{align*}

where $\beta$ is the approach to equilibrium for the rate limiting step. 

In the final expression, we note that:
\[ \beta =  \frac{\theta_\ce{O}\theta_\ce{CO}}{K_{2}\theta_\ce{CO2}} \]
\[ k_2 = v_2 e ^{-E_a/kT}\]
\[ K_1 = e ^{-\Delta G_1/kT}\]
\[ K_3 = e ^{-\Delta G_3/kT}\]
\[ K_4 = e ^{-\Delta G_4/kT}\]
where
$v_2$ is the attempt frequency, $k_2$ and $E_a$ are the rate constant and activation energy of the forward reaction for the RLS respectively, and $K_1$, $K_3$ and $K_4$ are the equilibrium constants of the corresponding steps.

The rate of reaction, in g/s is then given by:
\[ R_{mass} = M_{\ce{CO2}}*R_{overall}*A_s*N_s/N_A
\]
where $R_{mass}$, $M_{\ce{CO2}}$, $A_s$, $N_s$ and $N_A$ are the mass flow rate in g/s, molar mass of \ce{CO2}, number of sites available per area and the Avogradro number respectively. Solving these equations at fixed values of effective surface area $A_s = 2\times 10^7 $m$^2$, \ce{CO2} partial pressure of 2atm, a frequency factor of $5 \times 10^{13}$ and $\beta = 0.5$ yields the results reported in the paper. Varying these values over a wide range do not significantly affect the results since we are concerned with the ratio of the rates.

\bibliography{ref.bib}